\documentclass[12pt,preprint,showpacs,preprintnumbers,superscriptaddress,amsmath,amssymb,nofootinbib]{revtex4-1}

\usepackage{graphicx}
\usepackage{dcolumn}
\usepackage{bm,morefloats,color}

\catcode`\@=11
\@addtoreset{equation}{section}

\global\arraycolsep=1pt
\usepackage{amssymb,bm}

\def\Eq#1{\begin{equation} #1 \end{equation}}
\def\Eqr#1{\begin{eqnarray} #1 \end{eqnarray}}
\def\Eqrsubl#1#2{\begin{subequations}\label{#1}\Eqr{#2}\end{subequations}}

\newcommand{\nn}{\nonumber}
\newcommand{\pd}{\partial}

\newcommand{\bea}{\begin{eqnarray}}
\newcommand{\eea}{\end{eqnarray}}

\def\Xsp{{\rm X}}

\def\Ysp{{\rm Y}}
\def\Zsp{{\rm Z}}
\def\Wsp{{\rm W}}
\def\X5sp{{\rm X}_5}
\def\Y3sp{{\rm Y}_3}
\def\Z3sp{{\rm Z}_3}

\def\lap{{\triangle}}
\def\e{{\rm e}}

\begin{document}

\title{Probe brane dynamics on cosmological brane backgrounds}

\author{Kunihito Uzawa}
\affiliation{%
Department of Physics,
School of Science and Technology, 
Kwansei Gakuin University, Sanda, Hyogo 669-1337, Japan.
}%

\author{Kentaroh Yoshida}%
\affiliation{
Department of Physics, Kyoto University,   
Kyoto 606-8502, Japan. 
}%

\date{\today}

\begin{abstract}
We consider the dynamics of a single probe brane on various 
cosmological brane backgrounds. 
The on-shell condition of the static probe brane  
leads to the supersymmetric intersection rules for static BPS configurations, 
though the cosmological backgrounds do not preserve any supersymmetries. 
This is a remarkable feature associated with the cosmological backgrounds 
because in the static background the on-shell condition of the static brane 
gives no constraint on the brane configuration. 
Furthermore, it follows that under this condition
there is no velocity-independent force for the probe brane even on the cosmological backgrounds.   
\end{abstract}

\maketitle

\section{Introduction}
\label{sec:introduction}

String theory contains higher-dimensional $p$-branes $(p>1)$ as well as strings. 
The low-energy dynamics of the branes are described in supergravity theories. 
An innumerable number of brane solutions have been discovered so far. 
Most of them are static, but it is certain that cosmological brane solutions 
should also exist.  
For example, colliding brane solutions, which are found by Gibbons, Lu and Pope
\cite{Gibbons:2005rt},  are cosmological solutions 
(for the related progress, see \cite{Chen:2005jp, Kodama:2005fz, 
Kodama:2005cz, Minamitsuji:2010uz, Minamitsuji:2011jt, Minamitsuji:2012if, 
Uzawa:2013koa}). 
The colliding solutions have some potential applications in realistic cosmology 
\cite{Brill:1993tm, Maeda:2010aj, Maeda:2012xb, Uzawa:2013msa}. 
It may be related to a cosmic Big Bang of the origin of our Universe. 
Therefore, it is of great significance to understand the cosmological backgrounds 
profoundly. 
 
An interesting issue along this direction is to consider the brane dynamics on cosmological brane backgrounds. 
It is well studied on static brane backgrounds \cite{Tseytlin:1996hi}. The brane dynamics 
is an important subject in string theory and M-theory. For example, the matrix-model formulations 
of M-theory and type IIB string theory are intimately connected with the brane dynamics \cite{BFSS,IKKT}. 
It would be nice to consider cosmological backgrounds in the context of 
the matrix-model formulations. It would shed light on a new aspect of 
the non-perturbative formulation of superstring theories and M-theory and 
one may make a fundamental progress in string theory. 

In this manuscript, we consider the dynamics of a single probe brane on various 
cosmological brane backgrounds. 
This is formally a generalization of the work by Tseytlin \cite{Tseytlin:1996hi}\,, 
where ``{\it static}'' source brane backgrounds are considered. 
The on-shell condition of the static probe brane  
leads to the supersymmetric intersection rules for static BPS configurations, 
although the cosmological backgrounds do not preserve any supersymmetries. 
This is a remarkable feature associated with the cosmological backgrounds 
because in the static background the on-shell condition of the static brane 
gives no constraint on the brane configuration. 
Furthermore, it follows that under this condition
there is no velocity-independent force for the probe brane even on the cosmological backgrounds.   

This paper is organized as follows. Section \ref{sec:b} introduces cosmological brane backgrounds that 
we are concerned with here. 
In section \ref{sec:p}, we consider the dynamics of a single probe $p$-brane 
on the cosmological brane backgrounds. In particular, we argue on 
the condition 
under which the velocity-independent force vanishes. We will also study D-branes 
on the cosmological brane backgrounds in type IIB and IIA supergravities. 
Section \ref{sec:discussions} is devoted to conclusion and discussion. 

\section{Cosmological brane backgrounds}
\label{sec:b}

We first introduce cosmological brane backgrounds used in later discussion.  

The gravitational theory we are concerned with here includes the metric 
(in the Einstein frame) $\tilde{g}_{MN}$, a scalar field (dilaton) $\phi$, 
and a $(p+1)$-form field $A_{(p+1)}$ where the field strength is 
$F_{(p+2)}=dA_{(p+1)}$\,. 
This theory is realized by imposing some ansatz in type IIB and IIA supergravities. 
It is enough to describe charged $p$-branes.

The action is given by \cite{Lu:1995cs}
\Eq{
S=\frac{1}{2\kappa^2}\int \left[\tilde{R}\ast{\bf 1}_D
 -\frac{1}{2}\ast d\phi \wedge d\phi 
 -\frac{1}{2}\frac{\e^{\epsilon c\phi}}{(p+2)!}
 \ast F_{(p+2)}\wedge F_{(p+2)}\right]\,, 
\label{sb:action:Eq}
}
where $\kappa^2$ is the gravitational constant in $D$ dimensions and 
$\ast$ is the Hodge dual operator, $\tilde{R}$ denotes the Ricci scalar 
with the $D$-dimensional metric $\tilde{g}_{MN}$\,. Note that $\tilde{g}_{MN}$ 
is related to the metric $g_{MN}$ in the string frame via a Weyl rescaling:
\Eq{
g_{MN}=\e^{\phi/2}\,\tilde{g}_{MN}\,.
  \label{sb:cr:Eq}
} 
Then the constants $c$ and $\epsilon$ are defined as 
\Eqrsubl{sb:parameters:Eq}{
c^2& \equiv &4-\frac{2(p+1)(D-p-3)}{D-2}\,,
   \label{sb:c:Eq}\\
\epsilon& \equiv &\left\{
\begin{array}{cc}
 +&~~~~\mbox{for electric $p$-branes}\,,\\
 -&~~~~~~~\mbox{for magnetic $p$-branes}\,.
\end{array}\right.  
 \label{sb:epsilon:Eq}
   }
From the classical action (\ref{sb:action:Eq}), the field equations are obtained as 
\Eqrsubl{sb:fields:Eq}{
&&\hspace{-0.7cm}
\tilde{R}_{MN}=\frac{1}{2}\pd_M\phi \pd_N \phi 
+\frac{1}{2}\frac{\e^{\epsilon c\phi}}{(p+2)!}
\left[(p+2)F_{MA_2\cdots A_{p+2}} {F_N}^{A_2\cdots A_{p+2}}
-\frac{p+1}{D-2}\tilde{g}_{MN} F^2_{(p+2)}\right],
   \label{sb:Einstein:Eq}\\
&&\hspace{-0.1cm}\lap\phi-\frac{1}{2}\frac{\epsilon c}{(p+2)!}
\e^{\epsilon c\phi}F^2_{(p+2)}=0\,, 
   \label{sb:scalar:Eq}\\
&&\hspace{-0.1cm}d\left[\e^{\epsilon c\phi}\ast F_{(p+2)}\right]=0\,. 
   \label{sb:gauge:Eq}
}
Here $\tilde{R}_{MN}$ and $\lap$ are the Ricci tensor and the Laplacian, respectively, 
with respect to $\tilde{g}_{MN}$\,. 

Let us suppose the following form of the metric, 
\Eq{
d\tilde{s}^2=\tilde{g}_{MN}dx^Mdx^N=
\left[h(x, z)\right]^aq_{\mu\nu}(\Xsp)dx^{\mu}dx^{\nu} 
  +\left[h(x, z)\right]^bu_{ab}(\Zsp)dz^adz^b\,. 
 \label{sb:metric:Eq}
}
Here $x^M$ denotes the coordinates on the $D$-dimensional spacetime, 
X is a $(p+1)$-dimensional spacetime with the metric $q_{\mu\nu}$ and 
the coordinates $x^{\mu}$\,, and Z is a $(D-p-1)$-dimensional space with the metric $u_{ab}$ 
and the coordinates $z^a$\,. The parameters $a$ and $b$ are given by 
\Eq{
a=-\frac{D-p-3}{D-2}\,, \qquad b=\frac{p+1}{D-2}\,.
 \label{sb:parameter:Eq}
}
Then the parameter $c$ in \eqref{sb:c:Eq} can be rewritten as 
\Eq{
c^2=4\left[1-\frac{1}{2}ab(D-2)\right]\,.
}
The metric ansatz (\ref{sb:metric:Eq}) is a generalization of 
static $p$-branes with a dilaton coupling \cite{Lu:1995cs}. 
The cosmological brane solution can be obtained only in the particular case with (\ref{sb:parameter:Eq})\,, 
while the static brane solution does not need to satisfy the conditions in (\ref{sb:parameter:Eq})\,.

In addition, for $\phi$ and $F_{(p+2)}$\,, suppose the following forms, 
\Eqrsubl{sb:ansatz:Eq}{
&&\e^{\phi}=h^{\epsilon c/2}\,,
  \label{sb:phi:Eq}\\
&&F_{(p+2)}=d(h^{-1})\wedge\Omega(\Xsp)\,,
  \label{sb:f:Eq}
}
where $\Omega(\Xsp)$ is the volume $(p+1)$-form,
\Eq{
\Omega(\Xsp)=\sqrt{-q}\,dx^0\wedge dx^1\wedge \cdots \wedge dx^p\,, 
\qquad q \equiv \det q_{\mu\nu}\,.
}

Then the metric in (\ref{sb:metric:Eq}) should satisfy
\Eqrsubl{sb:equations:Eq}{
&&\hspace{-0.5cm} R_{\mu\nu}(\Xsp)=0\,, \qquad R_{ab}(\Zsp)=0\,,
   \label{sb:Ricci:Eq}\\
&&\hspace{-0.5cm} h(x, z)=h_0(x)+h_1(z)\,, \qquad 
D_{\mu}D_{\nu}h_0=0\,, 
\qquad \triangle_{\Zsp}h_1=0\,,
   \label{sb:h:Eq} 
 }
where $D_{\mu}$ is the covariant derivative with $q_{\mu\nu}$ and 
the Laplacian $\triangle_{\Zsp}$ is defined on the Z space. 
Similarly, $R_{\mu\nu}(\Xsp)$ and $R_{ab}(\Zsp)$ are the Ricci tensors
associated with $q_{\mu\nu}$ and $u_{ab}$, respectively. 

For later argument, we concentrate on a simple case specified with 
\[
q_{\mu\nu}=\eta_{\mu\nu}\,, \qquad u_{ab}=\delta_{ab}\,,
\] 
where $\eta_{\mu\nu}$ is the $(p-1)$-dimensional Minkowski metric, 
and $\delta_{ab}$ is the $(D-p-1)$-dimensional Euclidean metric.

Then the general solution of (\ref{sb:equations:Eq}) is given by 
\cite{Binetruy:2007tu, Maeda:2009zi}
\begin{eqnarray} 
h(x, z)= \left\{
\begin{array}{ll} \displaystyle 
\beta_{\mu}\,x^{\mu}+\bar{\beta}
+\sum_{l}\frac{M_l}{|z^a-z^a_l|^{D-p-3}} & \qquad (\mbox{for $D-p\ne 3$}) \\ 
\beta_{\mu}\,x^{\mu}+\bar{\beta}
+\sum_l M_l\ln |z^a-z^a_l|  &  \qquad (\mbox{for $D-p =3$}) 
 \label{sb:solution:Eq}
\end{array}
\right.\,,
\end{eqnarray}
where $\beta_{\mu}$, $\bar{\beta}$ and $M_l$ are real constants. 
The distance $|z^a-z^a_l|$ is defined as 
\[
|z^a-z^a_l|=\sqrt{\left(z^1-z^1_l\right)^2+\left(z^2-z^2_l\right)^2+\cdots+
\left(z^{D-p-1}-z^{D-p-1}_l\right)^2}\,. 
\]
When $\beta_0\neq 0$, the solution becomes cosmological. 
For $\beta_{\mu}=0$, the solution describes static BPS $p$-branes with 
charges $M_l$\,, which are aligned in parallel. 

In general, the dilaton does not vanish. There is no dilaton contribution 
in special cases with $c=0$\,, which contain   
\Eqrsubl{a:c=0:Eq}{
&&p=2~~~\mbox{and}~~~p=5~~~~~{\rm for}~~D=11\,, \nonumber \\
&&p=3~~~~~~~~~~~~~~~~~~~~~~~{\rm for}~~D=10\,. \nonumber 
}

In the following, we assume that $\beta_0 \neq 0$ and the other components are zero, 
for simplicity. 

\section{Probe branes on cosmological brane backgrounds} 
\label{sec:p}

We study the dynamics of a single probe $p$-brane on the cosmological brane backgrounds. 
First of all, as a simple case, we shall concentrate on a probe $p$-brane on cosmological 
$p$-brane backgrounds in $D$ dimensions with a constant dilaton or without a dilaton. 
It is shown that the on-shell condition of the probe brane leads to a constraint 
for the brane configuration. Then we consider a generalization to a probe $p_s$ brane 
on a cosmological $p_r$-brane background. 
Finally, we consider a D$p_s$-brane probe on a cosmological D$p_r$-brane background. 

\subsection{A probe $p$-brane on a cosmological $p$-brane 
background without dilaton}
\label{sub:warm}
We consider a probe $p$-brane on a cosmological $p$-brane 
background with the $(p+1)$-form $A_{(p+1)}$ in $D$ dimensions. 
For simplicity, we assume that the dilaton is constant or not contained. 
The analysis includes probe D3-branes in type IIB theory 
and M-branes in eleven dimensions. Then there is no distinction between 
the string frame and the Einstein frame. Furthermore, it is supposed that 
the NS-NS two-form and world-volume gauge field are turned off.  

In total, the probe $p$-brane action is simply given by 
\Eqr{
&&S_{p}=\int\!\! d\tau d^p\sigma\, {\cal L}
\left(\pd_{\tau}x^M\,,~\pd_{\alpha}x^M\right) =
-T_{p}\int d^{p+1}\sigma\,\sqrt{-\det \bar{g}_{\mu\nu}}
+T_{p}\int \bar{A}\,,
\label{m:action:Eq}
}
where $T_{p}$ is the $p$-brane tension. Then $\bar{g}_{\mu\nu}$ and $\bar{A}$ are 
the induced metric and $p$-form,  respectively,  
\Eq{
\bar{g}_{\mu\nu} \equiv g_{MN}\pd_{\mu}x^M\pd_{\nu}x^N\,, \qquad 
 \bar{A}_{\mu_1\cdots \mu_{p+1}}
=A_{M_1\cdots M_{p+1}}\pd_{\mu_1}x^{M_1}\cdots
\pd_{\mu_{p+1}}x^{M_{p+1}}\,. 
} 
The $p$-brane world-volume with the coordinates 
$\sigma^{\mu}~(\mu=0\,, \ldots\,, p)$ 
is embedded into a $D$-dimensional target spacetime 
via the functions $x^M(\tau\,, \sigma^{\alpha})~(M=0\,, \ldots\,, D-1)$\,. 

Let us argue a classical probe $p$-brane solution on a cosmological $p$-brane 
background with the metric (\ref{sb:metric:Eq})\,.  
We suppose that the probe $p$-brane is parallel to the background $p$-brane, 
for simplicity. 

We are interested in the following static configuration of the solution, 
\Eq{
t=\tau\,, \qquad x^{\alpha}=\sigma^{\alpha}\,, \qquad x^a =\mbox{const.}
\label{ap:gauge:Eq}
} 
By substituting (\ref{ap:gauge:Eq}) into the equation of motion, 
we obtain the following condition\footnote{Appropriate boundary conditions have to be  
imposed at the spatial infinity of the $p$-brane world-volume. 
Free endpoints are taken for all of the directions $x^M~(M=0\,, \ldots\,, D-1)$\,.}, 
\Eq{
\pd_t\left(h^{\frac{\xi}{2}}
-h^{-1}\right)=0\,.
\label{ap:equation2:Eq}
}
Here $\xi$ is defined as 
\Eq{
\xi \equiv -\frac{(p+1)(D-p-3)}{D-2}\,.
   \label{ap:xi:Eq}
}
For the cosmological backgrounds introduced in Sec.\,II, 
the constraint (\ref{ap:equation2:Eq}) is not satisfied in general, 
while it is trivially satisfied for the static backgrounds. The special case is $\xi=-2$\,. 
The configuration (\ref{ap:gauge:Eq}) becomes a classical solution if and only if $\xi=-2$\,. 
The condition $\xi=-2$ is realized for the following cases: 
\Eqrsubl{m:can:Eq}{
&&p=2~~~{\rm or}~~~5 \qquad~~ {\rm for}~~~D=11\,,\\
&&p=3  \qquad \quad\qquad\, ~~~{\rm for}~~~D=10\,.
}

Note that the first contribution in (\ref{ap:equation2:Eq}) comes 
from the Nambu-Goto part of the action and the second one from the coupling to the $p$-form gauge field. 
Thus, if the gauge-field contribution is ignorable 
(for example, when the probe $p$-brane is not parallel to the background $p$-brane),  
then $\xi=0$ is required as the on-shell condition.

The next task is to study the potential between the probe $p$-brane and 
the background $p$-branes. We assume that $\xi=-2$ so that the configuration (\ref{ap:gauge:Eq}) 
becomes a classical solution. In order to evaluate the potential, it is necessary to 
expand the original action (\ref{m:action:Eq})\,. 

First of all, let us expand the Nambu-Goto part. 
The metric in (\ref{sb:metric:Eq}) is diagonal and the determinant part 
in (\ref{m:action:Eq}) is expanded as 
\Eqr{
&&\sqrt{-\det\left(\bar{g}_{\mu\nu}\right)}
=\sqrt{-\det{g}_{\mu\nu}}\left[
1+\frac{1}{2}{g}^{\rho\sigma}g_{ab}\pd_{\rho}x^a\pd_{\sigma}x^b
+\frac{1}{8}\left({g}^{\rho\sigma}g_{ab}\pd_{\rho}x^a\pd_{\sigma}x^b\right)^2
\right.\nn\\
&&\left.\hspace{3cm}-\frac{1}{4}{g}^{\rho\sigma}\,{g}^{\alpha\beta}
\,g_{ab}\,g_{cd}\,
\pd_{\rho}x^a\pd_{\alpha}x^b\pd_{\sigma}x^c\pd_{\beta}x^d
+\cdots\right]\nn\\
&&~~~~=h^{\xi/2}\left[1+\frac{1}{2}
h\,\eta^{\rho\sigma}\,\delta_{ab}\pd_{\rho}x^a\pd_{\sigma}x^b\right.\nn\\
&&\left.~~~~~~~~~~~+\frac{1}{8}
h^{2}\,\eta^{\rho\sigma}\,\eta^{\alpha\beta}\,\delta_{ab}\,\delta_{cd}
\left(\pd_{\rho}x^a\pd_{\sigma}x^b\pd_{\alpha}x^c\pd_{\beta}x^d
-2\pd_{\rho}x^a\pd_{\alpha}x^b\pd_{\sigma}x^c\pd_{\beta}x^d\right)
+\cdots\right],
}
where ``$\cdots$'' denotes higher-order terms in derivatives. 
Note that the condition $\xi=-2$ indicates that the second-order terms in derivatives vanish and 
the higher-order terms start from the fourth order.   

Then the coupling term to the gauge-field in (\ref{m:action:Eq}) is expanded as 
\Eqr{
T_{p}\int \bar{A}&=&T_{p}\int \! d^{p+1}\sigma\,
\frac{1}{(p+1)!}\,\epsilon^{\nu_0\nu_1\cdots\nu_{p}}\,
\bar{A}_{\mu_0\mu_1\cdots\mu_{p}}
\frac{\pd x^{\mu_0}}{\pd\sigma^{\nu_0}}
\frac{\pd x^{\mu_1}}{\pd\sigma^{\nu_1}}\cdots
\frac{\pd x^{\mu_{p}}}{\pd\sigma^{\nu_{p}}}\nn\\
&=&T_{p}\int\! d^{p+1}x\,h^{-1}\,,
}
where we have used (\ref{sb:f:Eq}). 

In total, the original action (\ref{m:action:Eq}) is expanded as 
\Eqr{
\hspace{-0.5cm}S_{p}&=&T_{p}\int d^{p+1}x\left[-h^{\xi/2}\left\{1+\frac{1}{2} h 
\,\eta^{\rho\sigma}\,\delta_{ab}\pd_{\rho}x^a\pd_{\sigma}x^b\right.
\right.\nn\\
&&\left.\left.\hspace{-0.2cm}+\frac{1}{8}
h^{2}\,\eta^{\rho\sigma}\,\eta^{\alpha\beta}\,\delta_{ab}\,\delta_{cd}
\left(\pd_{\rho}x^a\pd_{\sigma}x^b\pd_{\alpha}x^c\pd_{\beta}x^d
-2\pd_{\rho}x^a\pd_{\alpha}x^b\pd_{\sigma}x^c\pd_{\beta}x^d\right)
+\cdots\right\}+h^{-1}\right]
   \label{m:action3:Eq} \\ 
   &=& T_{p}\int d^{p+1}x\left[ -h^{\xi/2} + h^{-1} \right] + \mbox{derivative terms}\,, \nonumber 
}
and the potential is obtained as 
\[
V = h^{\xi/2} - h^{-1}\,.
\]
Thus, for the on-shell condition $\xi=-2$\,,  the non-derivative corrections are canceled out 
and the potential starts at the fourth order in derivatives. This is the same result 
as in the static case \cite{Tseytlin:1996hi}. 
This indicates that the RR charge is equal to the tension of the probe $p$-brane.

When the probe $p$-brane is not parallel to the background $p$-branes, 
the potential (\ref{m:action:Eq}) receives the contribution only from the Nambu-Goto part 
because the coupling to $A_{\left(p+1\right)}$ vanishes. 
The on-shell condition leads to the condition $\xi=0$ and then the velocity-independent 
force vanishes. The condition $\xi=0$ implies $p=7$ (D7-brane) in $D=10$\,. 
However, the D7-brane background contains a non-trivial dilaton and hence 
this case is not included in the present analysis. A generalization including the dilaton 
is argued in Sec.\,\ref{sec:D}. 

\subsection{A probe $p_s$-brane on a cosmological $p_r$-brane background} 
\label{sec:int}

We next consider a probe $p_s$-brane on a cosmological $p_r$-brane background 
with a constant dilaton or without the scalar field.  

The classical $p_s$-brane action is given by 
\begin{eqnarray}
S_{p_s}=-T_{p_s}\int\! d^{p_s+1}\sigma\,\sqrt{-\det \bar{g}_{\mu\nu}}
+T_{p_s}\int \bar{A}
\equiv \int\!\! d\tau d^{p_s}\sigma\, {\cal L}
\left(\pd_{\tau}x^M,\pd_{\alpha}x^M\right)\,,
   \label{in:action:Eq}
\end{eqnarray}
where $T_{p_s}$ is the $p_s$-brane tension. Here $\bar{g}_{\mu\nu}$ and $\bar{A}$
are the induced metric and $p$-form, 
\begin{eqnarray}
\bar{g}_{\mu\nu}=g_{MN}\pd_{\mu}x^M\pd_{\nu}x^N\,, 
\quad 
\bar{A}_{\mu_1\cdots \mu_{p_s+1}}
=A_{M_1\cdots M_{p_s+1}}\pd_{\mu_1}x^{M_1}\cdots
\pd_{\mu_{p_s+1}}x^{M_{p_s+1}}\,. 
\end{eqnarray}

Suppose that the probe $p_s$-brane overlaps with the background 
$p_r$-branes in $\bar{p}$ spatial directions.  
Then it is suitable to rewrite the cosmological $p_r$-brane background as   
\Eqrsubl{in:solution2:Eq}{
&& ds^2=\left[h_r(t, z)\right]^{a_r}\left[\eta_{\mu\nu}(\Xsp)dx^{\mu}dx^{\nu}
+\delta_{ij}(\Ysp)dy^idy^j\right]\nn\\
&&~~~~~~+\left[h_r(t, z)\right]^{b_r}
\left[\delta_{mn}(\Wsp)dv^mdv^n
+\delta_{ab}(\Zsp)dz^adz^b\right], 
  \label{in:metric2:Eq}\\
&& A_{\left(p_r+1\right)}=h^{-1}_r(t, z)\,\Omega(\Xsp)\wedge 
\Omega(\Ysp)\,.
  \label{in:strength-r:Eq}
}
Here $\eta_{\mu\nu}$ is the $(\bar{p}+1)$-dimensional Minkowski metric 
and $\delta_{ij}(\Ysp)$ is the $(p_r-\bar{p})$-dimensional flat metric. 
Then $\delta_{mn}({\rm W})$ and $\delta_{ab}({\rm Z})$ are the $(p_s-p_r)$-dimensional 
and the $(9-p_s)$-dimensional flat metrics, respectively. 
The volume forms on the X and Y spaces are 
given by $\Omega(\Xsp)$ and $\Omega(\Ysp)$\,, respectively. 
The constants $a_r$ and $b_r$ are given by
\Eq{
a_r=-\frac{D-p_r-3}{D-2}\,,~~~~~b_r=\frac{p_r+1}{D-2}\,.
}
The background $p_r$-branes extend on the X and Y spaces, 
while the probe $p_s$-brane extends on the X and W spaces. 

Here we are interested in the following static configuration of the probe brane: 
\begin{eqnarray}
&&\sigma^0=x^0=t\,,\qquad 
\sigma^{\alpha}=x^{\alpha} \quad (\alpha=1\,,\cdots \,,\bar{p})\,, \nn\\
&&\sigma^{m}=v^{m} \quad (m=\bar{p}+1\,,\cdots \,,p_s)\,, \qquad
 x^a=\mbox{const.}
  \label{in:gauge3:Eq} 
\end{eqnarray}
Then the equation of motion leads to the following condition, 
\Eq{
\pd_th_r^{-\chi'/2}=0\,.
  \label{in:eq:Eq}
}
Here $\chi'$ is defined as 
\Eq{
\chi' \equiv \bar{p}+1-\frac{\left(p_r+1\right)\left(p_s+1\right)}{D-2}\,.
   \label{in:chi:Eq}
}
For the cosmological backgrounds, this condition is not automatically satisfied 
in comparison to the static backgrounds. 
Therefore the condition (\ref{in:eq:Eq}) is satisfied if and only if $\chi'=0$\,.  
That is, the on-shell condition leads to the condition $\chi'=0$\,. 

Let us evaluate the potential between the probe $p_s$-brane and 
the background $p_r$-branes. Similarly, 
the total action (\ref{in:action:Eq}) is expanded as 
\Eqr{
\hspace{-0.5cm}S_{p_s}&=&-T_{p_s}\int d^{p_s+1}x\,h_r^{\chi'/2}
\left[1+\frac{1}{2}h_r
\,\eta^{\rho\sigma}\,\delta_{ab}\pd_{\rho}x^a\pd_{\sigma}x^b
\right.\nn\\
&&\left.\hspace{-0.2cm}+\frac{1}{8}
h_r^{2}\,\eta^{\rho\sigma}\,\eta^{\alpha\beta}\,\delta_{ab}\,\delta_{cd}
\left(\pd_{\rho}x^a\pd_{\sigma}x^b\pd_{\alpha}x^c\pd_{\beta}x^d
-2\pd_{\rho}x^a\pd_{\alpha}x^b\pd_{\sigma}x^c\pd_{\beta}x^d\right)
+\cdots\right]
   \label{m:action4:Eq} \\ 
   &=& -T_{p_s}\int d^{p_s+1}x\,h_r^{\chi'/2}
   + \mbox{derivative terms}\,. \nonumber 
}
``$\cdots$'' denotes higher derivative terms. 
One can read off the potential $V$ as 
\Eq{
V(t, z)=h_r^{-\chi'/2}\,.
}
The on-shell condition means that $\chi'=0$ and then the velocity-independent force vanishes. 
Note that the derivative corrections start from the second-order. 

The condition $\chi'=0$ implies that the overlapping dimension 
$p$ is described as 
\Eq{
\bar{p}=\frac{\left(p_r+1\right)\left(p_s+1\right)}{D-2}-1\,.
  \label{in:no-force:Eq}
}
The relation (\ref{in:no-force:Eq}) is equivalent to the supersymmetric intersecting 
condition for the static branes when the dilaton is constant or not contained.  

In the eleven-dimensional supergravity, we get the intersections involving the
M2 and M5-branes 
\cite{Papadopoulos:1996uq, Strominger:1995ac, Townsend:1995af} 
(See also \cite{Maeda:2009zi, Minamitsuji:2010kb} 
for the dynamical brane background)
\Eq{
{\rm M2}\cap {\rm M2}=0\,,~~~~{\rm M2}\cap {\rm M5}=1\,,~~~~
{\rm M5}\cap {\rm M5}=3\,.
   \label{1:int M:Eq}
}

\subsection{A probe D$p_s$-brane on a cosmological D$p_r$-brane background}
\label{sec:D}

We study here a D$p_s$-brane on a cosmological D$p_r$-brane background.  
Note that the dilaton contribution is taken into account  (except for D3-branes). 
 
The classical D$p_s$-brane action is given by 
\Eqrsubl{pb:probe:Eq}{
S_{p_s} &=&-T_{p_s}\int d^{p_s+1}\sigma\,
\e^{-\phi}\sqrt{-\det\left(\bar{g}_{\mu\nu}+\mathcal{F}_{\mu\nu} 
\right)}
+T_{p_s}\int \bar{C}_{(p_s+1)}\,,
   \label{pb:action:Eq} \\
\bar{g}_{\mu\nu}&=&g_{MN}\pd_{\mu}x^M\pd_{\nu}x^N\,,  \quad  
\mathcal{F}_{\mu\nu} = \bar{B}_{\mu\nu}+2\pi\alpha'F_{\mu\nu}\,, \quad 
\bar{B}_{\mu\nu}= B_{MN}\pd_{\mu}x^M\pd_{\nu}x^N\,,\quad \\
\bar{C}_{\mu_1\cdots \mu_{p_s+1}}
&=&C_{M_1\cdots M_{p_s+1}}\pd_{\mu_1}x^{M_1}\cdots
\pd_{\mu_{p_s+1}}x^{M_{p_s+1}}\,.
}
Here $T_{p_s}$ is the D$p_s$-brane tension and $\bar{g}_{\mu\nu}$
is the induced metric. Then $\bar{B}_{\mu\nu}$ and $\bar{C}_{(p_s+1)}$ are 
the pullback of an NS-NS two-form and a $(p_s+1)$-form. 
The world-volume gauge field is given by $F_{\mu\nu}$\,. 
We work in the string frame hereafter. 

Note that the world-volume gauge-field strength $F_{\mu\nu}$ is turned off 
so that the probe brane does not carry the F-string charge, unless otherwise noted, 
because we are interested in the force between the probe brane without resolved F-strings 
and the background branes.

In the following, we will concentrate on two examples.
One is the case that 
a probe D$p_s$-brane is parallel to the background D$p_r$-branes.
The other is that a probe D$p_s$-brane overlaps with the background 
D$p_r$-branes in $\bar{p}$ directions.

\subsubsection{A probe D$p_s$-brane is parallel to the background 
D$p_r$-branes}

The background metric and fields are given by 
\Eqrsubl{rs:solution:Eq}{
ds^2&=&\left[h_r(t, z)\right]^{-1/2}\eta_{\mu\nu}(\Xsp)dx^{\mu}dx^{\nu}
+\left[h_r(t, z)\right]^{1/2}\delta_{ab}(\Zsp)dz^adz^b\,, 
 \label{rs:metric:Eq}\\
\e^{\phi}&=&h_r^{(3-p_r)/4}\,,
  \label{rs:dilaton:Eq}\\
C_{\left(p_r+1\right)}&=&h^{-1}_r(t, z)\,\Omega(\Xsp)\,\wedge \Omega(\Ysp)\,.
  \label{rs:strength-r:Eq}
}
Here $\eta_{\mu\nu}$ is the $(\bar{p}+1)$-dimensional Minkowski metric, 
and $\delta_{ab}$ is flat $(9-\bar{p})$-dimensional metric. 
Then $\Omega(\Xsp)$ denotes the volume form on the X space. 
A single probe D$p_s$-brane and the background D$p_r$-branes  
extend over the X space. 

For the static configuration, 
\Eqr{
&&\sigma^0=x^0=t\,, \qquad
\sigma^{\alpha}=x^{\alpha} \qquad (\alpha=1\,,\ldots \,,\bar{p})\,, \qquad
x^a=\mbox{const.}\,, 
 \label{rs:gauge:Eq}
}
the equation of motion for the probe D$p_s$-brane leads to the trivial 
condition, 
\Eq{
\pd_t\left(h_r^{-1}-h_r^{-1}\right)=0\,,
    \label{rs:equation:Eq}
}
and the resulting potential  
becomes constant. 
Thus there is no velocity-independent force. 

\subsubsection{A probe D$p_s$-brane overlaps with the background 
D$p_r$-branes}

The other case is that a single probe D$p_s$-brane overlaps with the background 
D$p_r$-brane in $\bar{p}$ directions. 
Then it is convenient to rewrite the background as  
\Eqrsubl{rs:solution2:Eq}{
ds^2&=&\left[h_r(t, z)\right]^{-1/2}\left[\eta_{\mu\nu}(\Xsp)dx^{\mu}dx^{\nu}
+\delta_{ij}(\Ysp)dy^idy^j\right]\nn\\
&& +\left[h_r(t, z)\right]^{1/2}
\left[\delta_{mn}(\Wsp)dv^mdv^n
+\delta_{ab}(\Zsp)dz^adz^b\right]\,, 
 \label{rs:metric2:Eq}\\
\e^{\phi}&=&h_r^{(3-p_r)/4}\,,
  \label{rs:dilaton2:Eq}\\
C_{\left(p_r+1\right)}&=&h^{-1}_r(t, z)\,\Omega(\Xsp)\,\wedge \Omega(\Ysp)\,. 
  \label{rs:strength-r2:Eq}
}
Here $\eta_{\mu\nu}$ is the $(\bar{p}+1)$-dimensional Minkowski metric. 
The flat metrics $\delta_{ij}$, $\delta_{mn}$ and $\delta_{ab}$ are defined in $(p_r-\bar{p})$\,, 
$(p_s-\bar{p})$ and $(9+\bar{p}-p_r-p_s)$ dimensions. Then 
$\Omega(\Xsp)$ and $\Omega(\Ysp)$ are the volume forms on the X space and the Y space, 
respectively. The background D$p_r$-branes extend on the X and Y spaces,  
while the probe D$p_s$-brane extends on the X and W spaces. 

For the static configuration, 
\Eqr{
&&\sigma^0=x^0=t\,, \qquad
\sigma^{\alpha}=x^{\alpha} \qquad (\alpha=1\,,\cdots \,,\bar{p})\,,\nn\\
&&\sigma^{m}=v^{m} \qquad (m=\bar{p}+1\,,\cdots \,,p_s)\,, \qquad
x^a=\mbox{const.}\,,
 \label{rs:gauge2:Eq}
}
the equation of motion for the probe D$p_s$-brane leads to the condition,   
\Eq{
\pd_t\left(h_r\right)^{(p_r+p_s-2\bar{p}-4)/4}=0\,.
    \label{rs:equation2:Eq}
}
This condition is satisfied if and only if  
\Eq{
p_r+p_s-2\bar{p}-4=0\,.
  \label{rs:no-force2:Eq}
}
Under this condition, the resulting potential $V$ is evaluated as 
\Eq{
V(t, z)=h_r^{(p_r+p_s-2\bar{p}-4)/4} = 1\,,
}
and hence there is no velocity-independent force.  
The condition in (\ref{rs:no-force2:Eq}) is equivalent to a supersymmetric intersection 
rule for the static D-branes \cite{Strominger:1995ac, Douglas:1995bn} 
(See also \cite{Binetruy:2007tu, Minamitsuji:2010kb} for 
the cosmological brane background)\,, 
\Eq{
{\rm D}p_r\cap{\rm D}p_r=\frac{1}{2}\left(p_r+p_s\right)-2\,.
  \label{10:rule:Eq}
}

\subsection{Comments on other cases}
Before closing this section, it is worth noting other cases. 
Our analysis is applicable to cases, a probe F-string and the background NS5-branes etc.  

Then the intersection rules \cite{Argurio:1998cp}\footnote{See also 
\cite{Minamitsuji:2010kb} for time dependent backgrounds.} 
involving F-string and NS5-brane are given by 
\Eqrsubl{1:int DNS:Eq}{
&&{\rm F1}\cap {\rm NS5}=1,~~~~
{\rm NS5}\cap {\rm NS5}=3,
   \label{1:int NS:Eq}\\
&&{\rm F1}\cap {\rm D}p=0,\\
&&{\rm D}p\cap {\rm NS}5=p-1,~~~~1\le p\le 6\,,
}
where $p$ is overlapping dimension of two branes. 
There is no solution for the F1-F1 and D0-NS5 intersecting systems
because the numbers of space dimensions for each pairwise overlap are negative by the intersection rule.

\section{Conclusion and Discussion}
  \label{sec:discussions}
  
We have considered the dynamics of a single probe brane on various 
cosmological brane backgrounds. 
The on-shell condition of the static probe brane  
leads to the supersymmetric intersection rules for static BPS configurations, 
although the cosmological backgrounds do not preserve any supersymmetries. 
This is a remarkable feature associated with the cosmological backgrounds 
because in the static background the on-shell condition of the static brane 
gives no constraint on the brane configuration. 
Furthermore, it follows that under this condition
there is no velocity-independent force for the probe brane even on the cosmological backgrounds.     

The dynamics of branes has continued to give a new insight 
in gravitational theories. Nowadays, it is of great interest 
to apply the brane dynamics to the construction of realistic cosmological models. 
Cosmological brane solutions would provide 
a strong bridge between string theory and cosmology. 
They could lead to realistic cosmological scenarios and 
then the brane dynamics would reveal the origin of the Universe. 

There remain some open problems such as a resolution of the curvature singularity 
in cosmological brane backgrounds. 
It is interesting to argue whether the singularity can be resolved 
at the string theory level  
by considering a probe brane moving on the cosmological brane background 
with the use of the present results. 
It would be an important key to understand the curvature singularity of a particular type, 
which may be related to the so-called enhancon mechanism that resolves a large class of spacetime 
singularities in string theory \cite{Johnson:1999qt, Jarv:2000zv}. 

In trying to construct cosmological models in string theory 
such as a brane inflation model \cite{Dvali:1998pa, Kachru:2003sx} 
in the early universe, probe branes are assumed in most of the models. Hence 
the dynamics of the probe brane on cosmological brane backgrounds  
would provide us a new tool to make the models more realistic. 
We hope to report progress in the near future.

\section*{Acknowledgments}
K.U. would like to thank M. Minamitsuji and 
T. Van Riet for discussions and valuable comments.


\end{document}